\documentclass[traditabstract,longauth,letter]{aa}
\usepackage{graphicx}
\usepackage{txfonts}
\usepackage{natbib}

\begin{document}

\newcommand{\mic}{\mbox{$\mu$m}}
\def\revision{ }

\title{The central region of spiral galaxies as seen by
  Herschel\thanks{Herschel is an ESA space observatory with science
    instruments provided by European-led Principal Investigator
    consortia and with significant participation from NASA.}}
\subtitle{M\,81, M\,99 \& M\,100}

\author{M. Sauvage\inst{\ref{inst11}},
N. Sacchi\inst{\ref{inst17}},
G. J. Bendo\inst{\ref{inst3}},
A. Boselli\inst{\ref{inst1}},
M. Pohlen\inst{\ref{inst2}},
C. D. Wilson\inst{\ref{inst20}},
R. Auld\inst{\ref{inst2}},
M. Baes\inst{\ref{inst5}},
M. J. Barlow\inst{\ref{inst6}},
J. J. Bock\inst{\ref{inst7}},
M. Bradford\inst{\ref{inst7}},
V. Buat\inst{\ref{inst1}},
N. Castro-Rodriguez\inst{\ref{inst8}}, 
P. Chanial\inst{\ref{inst11}},
S. Charlot\inst{\ref{inst9}},
L. Ciesla\inst{\ref{inst1}},
D. L. Clements\inst{\ref{inst3}},
A. Cooray\inst{\ref{inst25}},
D. Cormier\inst{\ref{inst11}},
L. Cortese\inst{\ref{inst2}},
J. I. Davies\inst{\ref{inst2}},
E. Dwek\inst{\ref{inst10}},
S. A. Eales\inst{\ref{inst2}},
D. Elbaz\inst{\ref{inst11}},
M. Galametz\inst{\ref{inst11}},
F. Galliano\inst{\ref{inst11}},
W. K. Gear\inst{\ref{inst2}},
J. Glenn\inst{\ref{inst13}},
H. L. Gomez\inst{\ref{inst2}},
M. Griffin\inst{\ref{inst2}},
S. Hony\inst{\ref{inst11}},
K. G. Isaak\inst{\ref{inst2},\ref{inst23}},
L. R. Levenson\inst{\ref{inst7}},
N. Lu\inst{\ref{inst7}},
S. C. Madden\inst{\ref{inst11}},
B. O'Halloran\inst{\ref{inst3}},
K. Okumura\inst{\ref{inst11}},
S. Oliver\inst{\ref{inst14}},
M. J. Page\inst{\ref{inst15}},
P. Panuzzo\inst{\ref{inst11}},
A. Papageorgiou\inst{\ref{inst2}},
T. J. Parkin\inst{\ref{inst20}},
I. Perez-Fournon\inst{\ref{inst8}},
N. Rangwala\inst{\ref{inst13}},
E. E. Rigby\inst{\ref{inst4}},
H. Roussel\inst{\ref{inst9}},
A. Rykala\inst{\ref{inst2}},
B. Schulz\inst{\ref{inst16}},
M. R. P. Schirm\inst{\ref{inst20}},
M. W. L. Smith\inst{\ref{inst2}},
L. Spinoglio\inst{\ref{inst17}},
J. A. Stevens\inst{\ref{inst18}},
S. Srinivasan\inst{\ref{inst9}},
M. Symeonidis\inst{\ref{inst15}},
M. Trichas\inst{\ref{inst3}},
M. Vaccari\inst{\ref{inst19}},
L. Vigroux\inst{\ref{inst9}},
H. Wozniak\inst{\ref{inst21}},
G. S. Wright\inst{\ref{inst24}},
W. W. Zeilinger\inst{\ref{inst22}}
}

\authorrunning{Sauvage, M. et al.}

\institute{	
CEA, Laboratoire AIM, Irfu/SAp, Orme des Merisiers, F-91191
Gif-sur-Yvette, France\label{inst11}
\email{marc.sauvage@cea.fr}
\and
Istituto di Fisica dello Spazio Interplanetario, INAF, Via del Fosso  
del Cavaliere 100, I-00133 Roma, Italy\label{inst17}
\and
Astrophysics Group, Imperial College, Blackett Laboratory, Prince  
Consort Road, London SW7 2AZ, UK\label{inst3}
\and
Laboratoire d'Astrophysique de Marseille, UMR6110 CNRS, 38 rue F.  
Joliot-Curie, F-13388 Marseille France\label{inst1}
\and
School of Physics and Astronomy, Cardiff University, Queens  
Buildings The Parade, Cardiff CF24 3AA, UK\label{inst2}
\and
Dept. of Physics \& Astronomy, McMaster University, Hamilton,  
Ontario, L8S 4M1, Canada\label{inst20}
\and
Sterrenkundig Observatorium, Universiteit Gent, Krijgslaan 281 S9,  
B-9000 Gent, Belgium\label{inst5}
\and
Department of Physics and Astronomy, University College London,  
Gower Street, London WC1E 6BT, UK\label{inst6}
\and
Jet Propulsion Laboratory, Pasadena, CA 91109, United States;  
Department of Astronomy, California Institute of Technology, Pasadena,  
CA 91125, USA\label{inst7}
\and
Instituto de Astrof\'isica de Canarias, v\'ia L\'actea S/N, E-38200 La  
Laguna, Spain\label{inst8}
\and
Institut d'Astrophysique de Paris, UMR7095 CNRS, Universit\'e Pierre  
\& Marie Curie, 98 bis Boulevard Arago, F-75014 Paris, France\label{inst9}
\and
Department of Physics \& Astronomy, University of California, Irvine,
CA 92697, USA\label{inst25}
\and	
Observational  Cosmology Lab, Code 665, NASA Goddard Space Flight   
Center Greenbelt, MD 20771, USA\label{inst10}
\and
Department of Astrophysical and Planetary Sciences, CASA CB-389,  
University of Colorado, Boulder, CO 80309, USA\label{inst13}
\and
ESA Astrophysics Missions Division, ESTEC, PO Box 299, 2200 AG
Noordwijk, The Netherlands\label{inst23} 
\and
Astronomy Centre, Department of Physics and Astronomy, University of  
Sussex, UK\label{inst14}
\and
Mullard Space Science Laboratory, University College London,  
Holmbury St Mary, Dorking, Surrey RH5 6NT, UK\label{inst15}
\and
School of Physics \& Astronomy, University of Nottingham, University  
Park, Nottingham NG7 2RD, UK\label{inst4}
\and
Infrared Processing and Analysis Center, California Institute of  
Technology, Mail Code 100-22, 770 South Wilson Av, Pasadena, CA 91125,  
USA\label{inst16}
\and
Centre for Astrophysics Research, Science and Technology Research  
Centre, University of Hertfordshire, College Lane, Herts AL10 9AB, UK\label{inst18}
\and
University of Padova, Department of Astronomy, Vicolo Osservatorio  
3, I-35122 Padova, Italy\label{inst19}
\and
Observatoire Astronomique de Strasbourg, UMR 7550 Universit\'e de  
Strasbourg - CNRS, 11, rue de l'Universit\'e, F-67000 Strasbourg\label{inst21}
\and
UK Astronomy Technology Center, Royal Observatory Edinburgh,
Edinburgh, EH9 3HJ, UK\label{inst24} 
\and
Institut f\"ur Astronomie, Universit\"at Wien, T\"urkenschanzstr. 17,  
A-1180 Wien, Austria\label{inst22}
}

\date{}

\abstract{ With appropriate spatial resolution, images of spiral
  galaxies in thermal infrared ($\sim$10\,\mic\ and beyond) often
  reveal a bright central component, distinct from the stellar bulge,
  superimposed on a disk with prominent spiral arms. {\em ISO} and
  {\em Spitzer} studies have shown that much of the scatter in the
  mid-infrared colors of spiral {\revision galaxies is related to changes}
  in the relative importance of these two components, rather than to
  other modifications, such as  the morphological type or star
  formation rate, that affect the properties of the galaxy as a
  whole. With the {\em Herschel} imaging capability from 70 to
  500\,\mic, we revisit this two-component approach at longer
  wavelengths, to see if it still provides a working description of the
  brightness distribution of galaxies, {\revision and to determine its
    implications on the interpretation of global far-infrared
    properties of galaxies.}

  We quantify the luminosity of the central component by both a
  decomposition of the radial surface brightness profile and a direct
  extraction in 2D. We find the central component contribution is
  variable {\revision within the three galaxies in our sample}, possibly
  connected more directly to the presence of a bar than to the
  morphological type. The central component's relative contribution is at its
  maximum in the mid-infrared range and drops around 160\,\mic\ to
  reach a constant value beyond 200\,\mic. The central component
  contains a greater fraction of hot dust than the disk component, and
  while the colors of the central components {\revision are} scattered,
  colors of the disk components are more homogenous from one galaxy to
  the next. }

\keywords{Infrared: galaxies -- Submillimeter: galaxies -- Galaxies:
  spiral -- Galaxies: nuclei -- Galaxies: fundamental parameters}

\maketitle

\section{Introduction}

The central region of spiral galaxies hosts a number of processes that
have little or no counterpart in the disks. A nuclear starburst,
triggered by bar instabilities and associated or not with an active
galactic nucleus (AGN) can often be found. The stellar density is also
enhanced with respect to the disk, leading to an increase in the
interstellar radiation field (ISRF) intensity. It is thus no surprise
that in the mid-infrared (MIR), spiral galaxies can often be seen as
the sum of a central component, hereafter referred to as the core
component as distinct from the bulge, and a disk component \citep[see
e.g.][]{bendo07}. Extraction of the core component from MIR brightness
profiles have revealed that it can contribute up to 90\% of the total MIR
luminosity and that the distribution of core/total luminosity ratio
is evenly spread between 0 and 1 \citep{rousselbar}. This study also
showed that much of the scatter in the MIR {\em global} properties of
spiral galaxies came from changes in the relative contribution of the
core component and of its spectral energy distribution (SED),
while disk components formed a much more homogenous group. In short,
the scatter affecting the {\em global} MIR colors of galaxies is
caused by a very {\em local} effect, i.e., the importance of the core
component, rather than by a change in any single property of galaxies such as the ISRF throughout the disk. Taking this into account,
\citet{rousselsfr} showed that the MIR luminosity of
galaxies could be turned into an accurate tracer of the star-formation
rate, provided it was corrected for the core contribution. More
recently, \citet{chanial} extended this analysis by showing that the
dispersion in the correlation between the infrared luminosity of
galaxies and the dust temperature could be reduced by considering 
a compactness parameter, again related to the relative
contribution of the core and disk components.

Care must therefore be taken when extrapolating locally established
relations, such as the correlation of star-formation tracers in
galactic disks, to global studies where galaxies are no longer
resolved. One first needs to verify that the total luminosity of
galaxies is indeed dominated by the component where these correlations
have been established, or to design methods to correct for the existence
of multiple contributors to the total luminosity. With its
unprecedented sensitivity and spatial resolution, {\em Herschel}
\citep{pilbratt} now allows us to carry this very investigation over
the full thermal infrared peak, by combining the PACS instrument
\citep{poglitsch}, delivering spatial resolutions of $5\farcs5$,
$6\farcs7$, and $11''$ (FWHM) at 70, 100, and 160\,\mic, and the SPIRE
instrument \citep{griffin} delivering spatial resolutions of
$18\farcs1$, $25\farcs2$, and $36\farcs9$ (FWHM) in its three
photometric bands at 250, 350, and 500\,\mic.

\section{The objects and data reduction methods}
\label{sec:sample}

Galaxies studied here are part of the Herschel Reference Survey
\citep{bosHRS}, a SPIRE volume-limited imaging survey of
323 galaxies. Out of the 4 noninteracting spirals observed during the
science demonstration phase, NGC\,3683 and NGC\,3982 are too small in
apparent size for the present study, leaving M\,99 and M\,100
\citep[see also][]{pohlen}, to which we add M\,81, which was observed as
part of the Very Nearby Galaxies program \citep{bendoSI}.
The properties of the objects that are relevant to this study are in Table~\ref{tab:sample}.

Details of the SPIRE data reduction can be found in \citet{pohlen} and
\citet{bendoSI}, while the PACS data reduction of M\,81 is described
in \citet{bendoSI}. Photometric uncertainties are quoted at $\pm$15\%
{\revision for all SPIRE bands and at $\pm$10/20\% for the PACS
  70/160\,\mic\ bands} by the instrument teams. We also use {\em
  Spitzer} MIPS maps at 24, 70, and 160\,\mic\ \citep[{\revision see}][{\revision
  for data reduction details}]{bendoMips, clem10}. The SPIRE maps are
calibrated in Jy/Beam, and we used a beam size of
9.28$\times10^{-9}$, 1.74$\times10^{-8}$, and 3.57$\times10^{-8}$\,sr
at 250, 350, and 500\,\mic, respectively, to convert these into more
conventional brightness units. The MIPS maps are calibrated in MJy/sr,
and the PACS maps in Jy/pixel, with a pixel size of 1$\farcs$4 and
2$\farcs$85 for the 70 and 160\,\mic\ bands.

\begin{table}
  \caption{Relevant data for the galaxies {\revision studied} here, as well as the images used (M: MIPS, P: PACS, S: SPIRE).}
\label{tab:sample}
\centering
\begin{tabular}{lllll}
  \hline
  Name & Type & Nuc$^a$  & $D_{25}$ ($'$) & available data  \\
  \hline
  M\,81 & SAab      & Sy\,1.5 & 26.92 & M(24, 70, 160\,\mic),  \\
  &                 &              &            & P(70, 160\,\mic), \\
  &                 &              &            & S(250, 350, 500\,\mic) \\
  M\,99 & SAc         & HII        & 5.37   &  M(24, 70, 160\,\mic) \\
  &                 &              &            & S(250, 350, 500\,\mic) \\
  M\,100 & SABbc & T2         & 7.41   & M(24, 70, 160\,\mic) \\
  &                 &              &            & S(250, 350, 500\,\mic) \\
  \hline
\end{tabular}

\begin{list}{}{}
\item[$^{a}$] {Nuclear classifications from \citet{ho97}, T2 meaning a
    transition object between a starburst and a Sy\,2.}
\end{list}

\end{table}

\section{Extraction of the core components}
\label{sec:analysis}

Our first approach uses radial surface brightness profiles measured
inside concentric ellipses. For all maps of a given galaxy, we 
used the same parameters to build the integrating ellipses, namely the
central coordinate, the ellipse axis ratio, and the position angle;
i.e., these parameters are not fit to the data but obtained from
NED\footnote{{\tt http://nedwww.ipac.caltech.edu/}\,.}. This is a
coarser approach than the one used by \cite{pohlen}, where the ellipse
parameters are first fitted to the maps before the profile integration
is made. A simpler approach is justified here since we are not searching
for the most accurate structural description of the surface brightness
distribution in galaxies, but rather for a method to systematically
measure the contribution of the central region to the total galaxy
luminosity. Errors in the computed average surface brightness are the
quadratic sum of the error on the mean brightness, of the error
derived by accounting for the corresponding error map on the same
ellipses, and of the error on the background level. Generally, the total
errors are much smaller than the computed brightness except close to
the background. {\revision For all maps, we intentionally went significantly
  beyond the $D_{25}$ diameter.} To obtain the galaxy's brightness
profile, we determined the background brightness as the mean of the
brightness measured typically about 100$"$ beyond the point where the
galaxy emission ceased to be detected.

To search for and extract the core component, we obtained the best (in
the least-square sense) combination of {\revision Gaussian} or exponential
functions that represent the measured radial profiles. Expressed
in units of surface brightness as a function of radius, we fit the
following function to the data:
\begin{equation}
S(r) = \sum_{i=1}^{N} S^{0}_{i} \exp - \left(\frac{r}{r^{s}_{i}}\right)^{k_{i}},
\end{equation}
with $S^{0}$ the central surface brightness, $r^{s}$ the scale length,
$k=1,2$ for exponential or {\revision Gaussian} functions\footnote{in the
  case of a {\revision {\revision Gaussian}}, we have $r^{s}=0.6 \times $FWHM.},
and $N=$1 or 2. The fitting function was smoothed by the appropriate
PSF, so the parameters we derived describe the intrinsic light
distribution. It is important to recall that our aim is to build a
robust and systematic method to get at the one parameter we are
searching for: the relative contribution of the core component
to the total flux, and not to achieve the best representation of the full
profile.

The use of either {\revision Gaussian} or exponential functions is a result
of the fit, but that some of the disk profiles prefer a {\revision
  Gaussian} to an exponential must not be overinterpreted because the
presence of prominent spiral arms can create a distinctive bump in the
disk part of the profile that is represented by a {\revision
  Gaussian} better than by an exponential. In either case we did not use the fit
parameters that describe the ``disk'' component in the rest of the
analysis. The central component is almost always represented
by a {\revision Gaussian} profile very well, with the disk component also adequately fit 
at the location where the core profile merges into the disk
profile (see Figure~\ref{fig:prof}). We then decomposed the luminosity
of the galaxies in the following way. The total luminosity was obtained
by integrating the measured radial profile up to the background
radius, the total luminosity of the core component was obtained by
integrating the fitted profile of the core, and the disk luminosity is
simply the difference between these two quantities. In
Table~\ref{tab:results} we list the core components' scalelength and
contribution to the total luminosity derived from the fits.

In a second approach, we performed 2D fitting on M81 (chosen because
of the level of details available and the existence of a point-like
component in the center), using the {\tt GALFIT} code \citep{peng}.
Galactic structure (arms, bars, rings) is much more complex to
represent mathematically in 2D, thus this method converges on the
extraction of unresolved components. This provides a cross-check to
the radial profile approach since, when the central component is
resolved in the profile, the 2D method should underestimate it, but both methods should agree
when it is unresolved. This is what we
observe (see below).

\begin{table}
\caption{Scale-lengths and relative contributions of the core component. }
\label{tab:results}
\centering
\begin{tabular}{|l|r|l|r||r|l|r|}
\hline
& \multicolumn{3}{|c||}{M\,81} & \multicolumn{3}{|c|}{M\,100}\\
\hline
Band & $r^{s}$ ($''$) & Fit & $f_{c}$ & $r^{s}$ ($''$) & Fit & $f_{c}$ \\
(1) & (2) & (3) & (4) & (5) & (6) & (7) \\
\hline
M\,24\,\mic & 7.2 & E+E & 8.8  & 10.3 & G+G & 20.1 \\
M\,70\,\mic & 38.8 & G+G & 6.2 & 13.2 & G+G & 12.2 \\
P\,70\,\mic & 27.9  & G+G  & 6.3 & \multicolumn{3}{c|}{{\em n.a.}} \\
M\,160\,\mic & 28.5& G+G & 0.3 &  5.14 & G+G & 3.5\\
P\,160\,\mic &  26.4 & G+G  & 1.2 & \multicolumn{3}{c|}{{\em n.a.}} \\
S\,250\,\mic & 25.2 & G+G & 0.3 &  9.5 & G+G & 7.8 \\
S\,350\,\mic & 10.2 & G+G & 0.2 &  5.4 & G+G & 7.4 \\
S\,500\,\mic & 11.8 & G+G & 0.2 &  6.5 & G+G & 7.6\\
\hline
\end{tabular}

\begin{list}{}{}
\item[] Notes.-- For M\,99, the profile fit revealed no central component except at 24\,\mic\ so we report this in the text. In column (1), M, P and S are MIPS, PACS and SPIRE, respectively, (2) and (5) the core component scale-length in arcseconds, (3) and (6) the best fit functions, E for exponential and G for {\revision Gaussian}, and (4) and (7) the core component contribution to the total flux in \%, respectively for M\,81 and M\,100. Relative uncertainties on the scale parameters are typically of $10\%$, while on the core fractions these are typically $20\%$
\end{list}
\end{table}

\section{Discussion}
\label{sec:results}

\begin{figure}
\hbox{
\includegraphics[width=8.8cm]{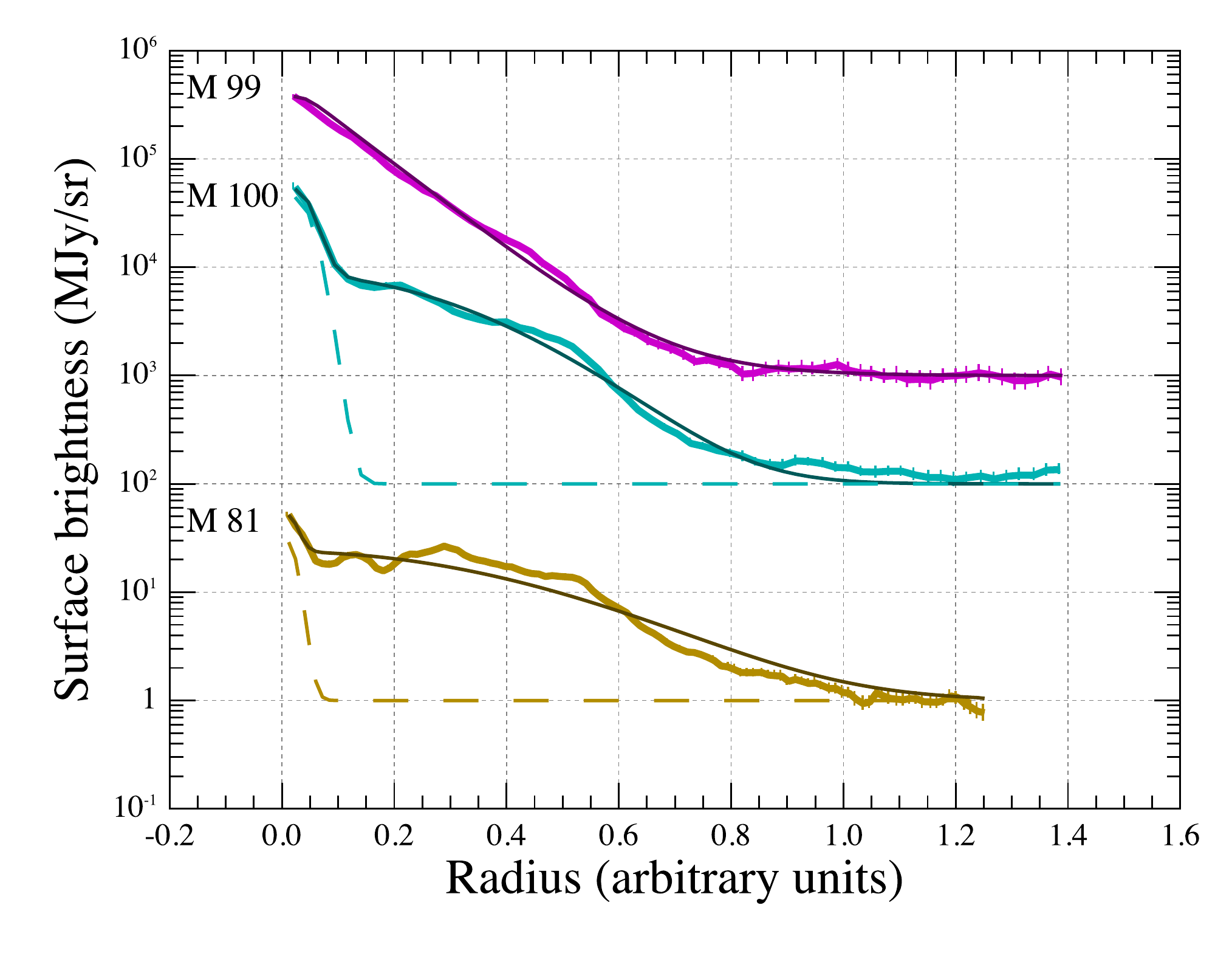}
}
\caption{The surface brightness radial profiles at 250\,\mic\ (thick curves with error bars) and the fitted profiles. We also plot as dashed lines the core component fit, to visualize its relative contribution. A value of 1 was added to all profiles to avoid visually enhancing the noise in the outer part of the profile, and the M\,99 and M\,100 profiles are multiplied by $10^2$ and $10^3$ for clarity. The abscissa of the graph is the radius normalized by the distance at which we measure the background.  }
\label{fig:prof}
\end{figure}

In general, we observe that the same family of functions provides the best-fit models to a given galaxy, regardless of the wavelength. This is mostly because the same morphological features dominate the profile of galaxies over the explored spectral domain, a fact worth noting given the quite drastic changes in the thermodynamical state and physical nature of the mid to far IR-emitting grains, from out-of-equilibrium, nearly molecular-sized grains to thermalized sub-micron grains \citep[but see][for a discussion of the full profiles properties]{pohlen}. However, the profiles already show that the relative importance of the core component is variable from one galaxy to the next (see Figure~\ref{fig:prof}). 

{\em M\,81 -} This is the closest galaxy and thus its central region
is observable with a high level of detail \citep[see][]{bendoSI}. The core is compact, possibly the IR counterpart of the Seyfert
nucleus, on which a bright arc, presumably the start of one of the
spiral arms, connects. Moving away from the center, the brightness of
the arm diminishes until it increases again to form a bright, almost
ring-like structure. Thus all profiles, except the 24\,\mic\ one, show
a first local minimum at $\sim80''$ radius (1.4\,kpc at 3.65\,Mpc).
With high enough resolution, the double nature of the core component
(central {\revision unresolved} source and spiral arm) is visible in the
profile as a break in the slope of the core profile. The interplay
between the resolved and unresolved parts of the core is clear in the
comparison of the {\tt GALFIT} and profile-fitting {\revision core-fraction
  estimates}: as the former preferentially extracts the unresolved
component, while the latter {\revision is able to} extract both the
unresolved and {\revision resolved} components as a single {\revision structure}
departing from the disk, {\revision we expect both methods to agree, for a
  given band, when either the unresolved source is dominant or the
  whole core structure is unresolved. This is indeed observed for MIPS
  24\,\mic\ (where the unresolved component dominates) and for MIPS
  160, SPIRE 350, and 500\,\mic\ (where the whole core structure is not,
  or barely, resolved)}. In the other bands, the unresolved component is
less dominant, and {\tt GALFIT} computes core fractions that are a
factor of 3 below those coming from the profile fitting. However,
inspection of the corresponding images clearly shows the unresolved
component sitting on a core structure distinct in brightness from the
disk. The core contribution derived with the profile fitting method is
also very compatible with what can be measured naively from the
profile itself, inside the 80$''$ radius. The core component is
generally well represented by {\revision Gaussians} and, when resolved, the
derived scalelengths are compatible with one another, given
typical uncertainties of 10\%. We note a severe disagreement between the
two 160\,\mic\ core fraction measurements, but inspection of the
profiles clearly shows that this stems from dilution of the barely
resolved central component in the disk emission at the low resolution
of {\em Spitzer}, thereby pointing to the intrinsic limitations of the
MIPS\,160\,\mic\ band as the way to perform these decomposition studies. At
increasing SPIRE wavelengths, we observe as well that the central
component becomes gradually unresolved.

{\em M\,99 -} Although this galaxy clearly shows a bright center and
spiral arms in all available IR bands \citep[see][]{pohlen}, our fit
of the profiles is usually adequate with a single exponential
function; however: (1) an exponential does not accurately represent
the disk profile that has a distinct break \citep[at $r\simeq0.4$ in
Figure~\ref{fig:prof}, see][for a discussion of this feature]{pohlen},
and (2) the 24\,\mic\ image is sharp enough to reveal a barely resolved {\revision Gaussian} component in the center
($r^{s}=3\farcs1$) contributing less than 0.5\% of the total 24\,\mic\
flux.

{\em M\,100 -} The barred nature of the galaxy clearly shows up as a bump
in the high-resolution profiles \citep[e.g. 24\,\mic\ and 250\,\mic,
see also][]{pohlen}, and as a result, the fitting function favors a
double {\revision Gaussian} profile, with very good fits of the central
component achieved from 24\,\mic\ to 500\,\mic.
Table~\ref{tab:results}, however, shows the central component is  unresolved at 160, 350, and 500\,\mic, while for the other wavelengths a
scale-length of $11''\pm2''$ is indicated by the fits. When the
central component is unresolved, our capacity to accurately measure
its contribution will depend on how bright it is compared to the disk
component. Thus it is possible that the core fraction we compute at
160\,\mic\ (the band with the worst resolution) is in fact a lower
limit to the actual core fraction, pointing again to the limited
ability of the MIPS 160\,\mic\ band to accurately measure core
contributions.

\begin{figure}
\includegraphics[width=8.8cm]{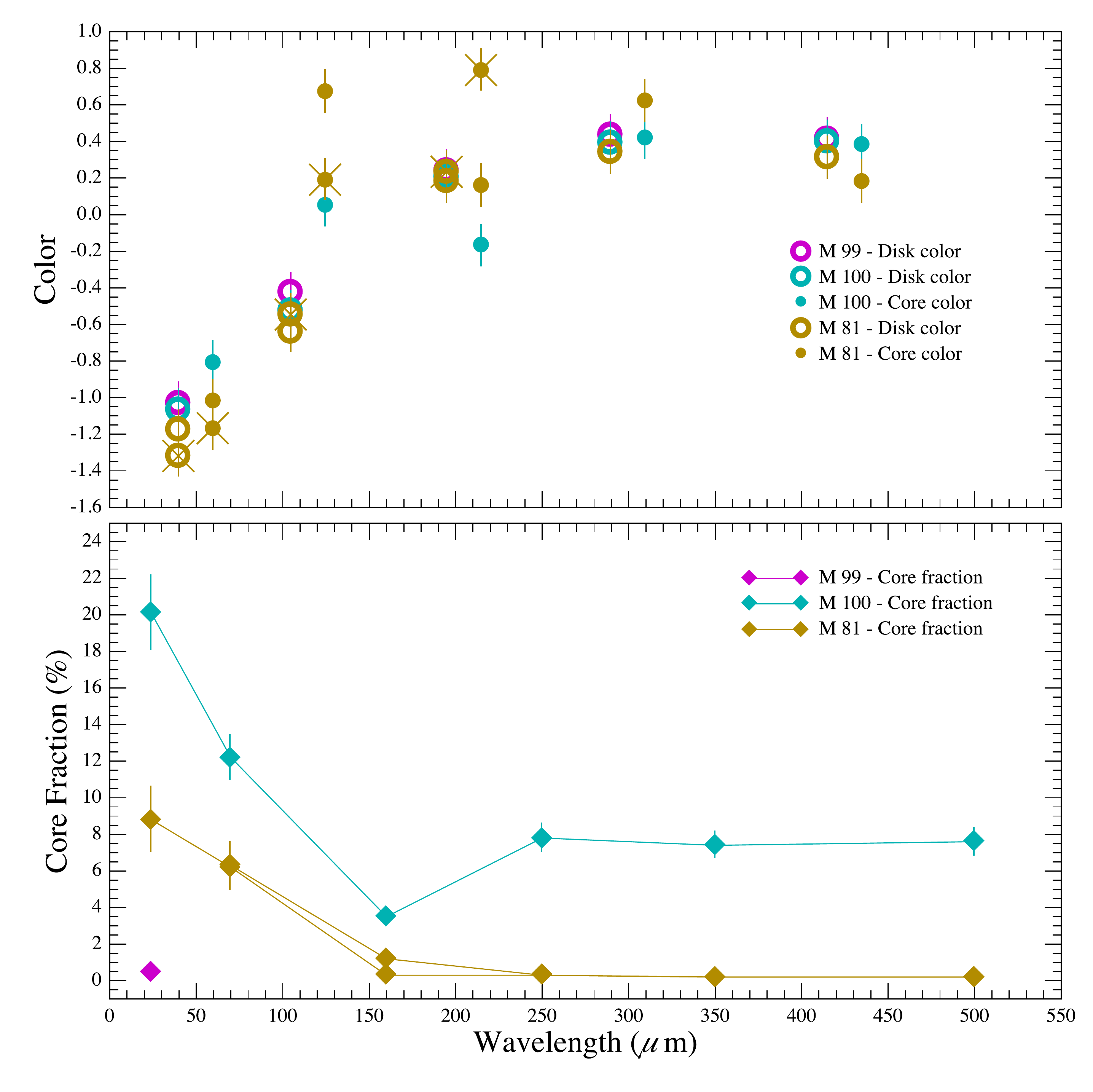}
\caption{Top panel: colors as a function of wavelength. Colors are displayed at an abscissa in-between the band wavelengths used, e.g., the [24]/[70] color is plotted at 50\,\mic, with open symbols for the disk colors and filled symbols for the core colors. For M\,81, we identify the colors incorporating PACS data with an extra cross overlaid on the corresponding symbol. For clarity, the core and disk colors are offset in x by 10\,\mic. Bottom panel: Core fractions as a function of wavelength, again with two series of data for M\,81, with MIPS and PACS. Error bars can be smaller than symbol size.}
\label{fig:corcol}
\end{figure}

The necessity to account for a core component to represent the surface
brightness profile already appears weakly related to the
morphological type: it is on the intermediate type galaxy M\,100 that
we observe the largest core fraction. It is quite tempting, and
probably logical, to relate that to the presence of a weak bar in
M\,100, but this can only be asserted on a reasonable sample, as the presence of a bar does not
guarantee there is a strong MIR core component
\citep{rousselbar}.

Interestingly, even though M\,81 and M\,100 have significantly
different core fractions (by a factors $>$2), we observe in
Figure~\ref{fig:corcol} the same trend in the evolution of this
fraction with wavelength: it is maximum at the shorter wavelengths
and nearly constant beyond 200\,\mic, with the transition occurring
in the 70-200\,\mic\ range. The evolution of the core fraction can be
interpreted by inspecting the component's colors. We show in
Figure~\ref{fig:corcol} the colors of the core and disk components
built from neighboring bands. This figure first shows quite clearly
that all the disk colors are very similar for the
three galaxies studied here, even when the core component represents a
substantial fraction of the emission (10-20\% at
$\lambda\leq160\,\mic$). It also shows that only the
[70]/[160]\footnote{shorthand for $\log
  \left[f_{\nu}(70\mic)/f_{\nu}(160\mic)\right]$.} core colors, and to
a lesser extent [160]/[250], significantly differ from the disk
colors. The [70]/[160] core colors are systematically higher than
those of disks, showing the core emission as coming from hotter dust
than the disk emission. If the dust is hotter in the core than in the
disk, then it will represent a larger fraction of the emission at
shorter wavelengths. In the SPIRE range, we reach the Rayleigh-Jeans
regime, where colors have no dependence on temperature. That
color differences are more prominent in the 70-250\,\mic\ range is
simply because this is where the peak of the thermal emission of dust
in galaxies occur, showing that this range is key to properly
understanding the infrared emission of galaxies.

To reconcile the small [24]/[70] color variation between disks and
cores with the larger core fractions at 24\,\mic, we suggest that it
is because the 24\,\mic\ and some of the 70\,\mic\
emission come from transiently heated grains, i.e., show little
spectral variation with the ISRF intensity or hardness. A harder ISRF
in the core could then explain the larger core fractions observed at
24\,\mic\ by boosting the emission from transiently heated grains
without significantly affecting their emission spectrum. The small AGN
in M\,81 is a natural explanation for this, as is a possible
bar-driven starburst in the center of M\,100. We also tested an
alternative explanation in M81, namely a direct contribution of bulge
stars in the 24\,\mic\, band, as suggested by \citet{bendo08}, by
correcting the MIPS image using IRAC data {\revision (a correction to the
  total flux of $-5\%$). The ``dust-only'' core fraction is almost
  unchanged, revealing that bulge stars are not responsible for the FIR
  core structure}.

\section{Conclusions}
\label{sec:conc}

In this small set of galaxies, the core fractions are small
($\leq{20}\%$), therefore the correction to the full SED of the
galaxies introduced by accounting for the different core properties
would be minimum. Nevertheless we already observe both a significant
variation in the core component contribution from one galaxy to the
next and a systematic decrease with wavelength. We also find
that the galactic disks studied here show remarkable color homogeneity
over the 24-500\,\mic\ range, while the core components are notably
hotter as shown by their higher [70]/[160] color value. This opens the
possibilities that (i) across the {\em Herschel} wavelength range, the
global SED of a galaxy may be influenced as much by the relative
importance of the core and disk components as by a global property,
such as the star formation rate or the morphological type, and (ii) that
this may depend on the observed wavelength. We shall thus examine this
in larger samples such as the HRS set of resolved spiral galaxies.

\acknowledgements{ SPIRE has been developed by a consortium of
  institutes led by Cardiff University (UK) and including Univ.
  Lethbridge (Canada); NAOC (China); CEA, OAMP (France); IFSI, Univ.
  Padua (Italy); IAC (Spain); Stockholm Observatory (Sweden); Imperial
  College London, RAL, UCL-MSSL, UKATC, Univ. Sussex (UK); and
  Caltech/JPL, IPAC, Univ. Colorado (USA). This development has been
  supported by national funding agencies: CSA (Canada); NAOC (China);
  CEA, CNES, CNRS (France); ASI (Italy); MCINN (Spain); Stockholm
  Observatory (Sweden); STFC (UK); and NASA (USA).

  PACS has been developed by a consortium of institutes led by MPE
  (Germany) and including UVIE (Austria); KUL, CSL, IMEC (Belgium);
  CEA, OAMP (France); MPIA (Germany); IFSI, OAP/AOT, OAA/CAISMI, LENS,
  SISSA (Italy); IAC (Spain). This development has been supported by
  the funding agencies BMVIT (Austria), ESA-PRODEX (Belgium), CEA/CNES
  (France), DLR (Germany), ASI (Italy), and CICT/MCT (Spain).

  This research has made use of HIPE, a joint development by the
  Herschel Science Ground Segment Consortium, consisting of ESA, the
  NASA Herschel Science Center, and the HIFI, PACS and SPIRE
  consortia, and of the NASA/IPAC Extragalactic Database (NED) which
  is operated by the Jet Propulsion Laboratory, Caltech, under contract with NASA. }

\end{document}